\documentstyle[aps,epsf]{revtex}
\def\be{\begin{equation}}
\def\ee{\end{equation}}
\def\lan{\langle}
\def\ran{\rangle}
\def\tp{t^\prime}
\draft
\title{Spin and density overlaps in the frustrated Ising lattice gas}
\author{Antonio de Candia and Antonio Coniglio}
\address{Dipartimento di Scienze Fisiche\\
INFM, Unit\`a di Napoli\\
Monte Sant'Angelo, I-80126 Napoli, Italy}
\begin{document}
\twocolumn
\wideabs{%
\maketitle
\begin{abstract}
We perform large scale simulations of the frustrated
Ising lattice gas, a three-dimensional lattice model of a structural
glass, using the parallel tempering technique. We evaluate
the spin and density overlap distributions, and the corresponding non-linear
susceptibilities, as a function of the chemical potential.
We then evaluate the relaxation functions of the spin and density self-overlap,
and study the behavior of the relaxation times.
The results suggest that the spin variables undergo a transition very similar
to the one of the Ising spin glass, while the density variables do not 
show any sign of transition at the same chemical potential. It may be that
the density variables  undergo a transition at a higher chemical potential,
inside the phase where the spins are frozen.
\end{abstract}
\pacs{}
}
%

\section{Introduction}
In the last years, a scenario has emerged for the theoretical
description of structural glasses, in which the behavior of glasses is
characterized by two different temperatures. The higher temperature $T_c$,
identified  with the critical temperature of the ideal mode-coupling
theory\cite{ref:mct},
corresponds to a crossover in the dynamical behavior of the glass,
characterized by a single relaxation
time for $T>T_c$, and by a two-step relaxation for $T<T_c$.
Below $T_c$, the free
energy landscape in the phase space is splitted into an extensive number of     valleys, separated by high barriers, and the dynamics is separated in
a fast motion inside the valley, and a much slower motion among different
valleys.
The number of valleys accessible to the system at the temperature $T$,
can be expressed by the formula
${\cal N}=\exp(N\Sigma(T))$, where $N$ is the number of particles in
the system, and $\Sigma(T)$ is the so-called configurational entropy.
The second temperature $T_K$, identified with the Kauzmann
temperature\cite{ref:kauzmann}
and the critical temperature of the Adam and Gibbs theory\cite{ref:adams},
is characterized by the vanishing of the
configurational entropy $\Sigma(T)$, and the divergence of the relaxation
time of the system with a Vogel-Fulcher-Tamman law.

This picture is corroborated by the analogy with a class of mean field
models, the $p$-spin glasses. 
%
%
These models undergo a dynamical
transition at a temperature $T_d$, where the phase space splits
into an extensive number of metastable states, and the correlation functions
show a singularity of the same type of the one predicted by mode-coupling
theory\cite{ref:kirks}.
At a lower temperature $T_s$ the number of metastable states becomes 
non-extensive, that is the configurational entropy vanishes, and for $T<T_s$
the model shows a 1-step replica symmetry breaking.
%
%
It has been longly debated to what extent this analogy, between
structural glasses and $p$-spin models, can be pushed forward. 
In particular one may ask if structural glasses exhibit some kind of
replica symmetry breaking, and if they are in the same universality class
of $p$-spin models or not. There are presently some results that point
in this direction.
First principle computations of the equilibrium thermodynamics
of simple fragile glasses seem to be consistent with this picture
\cite{ref:parimez},
and the off-equilibrium fluctuation-dissipation
ratio, in molecular dynamics simulations of supercooled liquids
\cite{ref:fdt}, shows the
same pattern observed in $p$-spin glasses and predicted for 1-step RSB
models\cite{ref:kurch,ref:peliti}.

Nevertheless, there are still some important differences between structural
glasses and $p$-spin models.
In $p$-spin glasses
the relevant variables are the spin orientations, and there is explicit
quenched disorder in the Hamiltonian.
On the other hand, in glasses the relevant variables are the
particle positions, and there are no quenched interactions in the Hamiltonian,
but the disorder is ``self-generated'' by the geometrical hindrance that the
particles exert on each other.
Furthermore, $p$-spin glasses are mean-field models, while structural glasses
live in finite dimension. This means among other things that metastable states
in $p$-spin models have infinite lifetime, and the dynamical transition
$T_d$ corresponds to a divergence of the relaxation times, while in finite
dimensional models the barriers between free energy valleys can be
overcome by ``hopping'' processes, which restore ergodicity
even below the crossover temperature $T_c$.
Furthermore the $p$-spin model, if studied in three 
dimensions, loses many of its mean field ``glassy'' properties,
and shows a transition similar to the full replica symmetry breaking
transition of Ising spin glasses, though with some
remarkable differences\cite{ref:3dps}.

The frustrated Ising lattice gas was introduced some time ago
as a simple finite dimensional lattice model
of a glass-forming liquid\cite{ref:filgmc}, in order to overcome some
of the limitations of $p$-spin models.
Each lattice site
carries two kinds of variables, a lattice gas variable $n_i=0,1$,
which represent
the presence or absence of a particle on the $i$-th site, and an Ising
spin variable $S_i$, which represent an internal degree of freedom of the
particle, such as for example the orientation of a non-symmetrical molecule.
The Hamiltonian of the model is
\be
{\cal H} = J\sum_{\lan ij \ran}
(1-\epsilon_{ij}S_i S_j)n_in_j -\mu \sum_i n_i,
\label{eq:filg}
\ee
where $\epsilon_{ij}=\pm 1$ are quenched variables.
In the limit $J\to\infty$, the first term of the Hamiltonian
implies that two nearest neighbor sites can be
simultaneously occupied by two particles only if their spin variables
satisfy the constraint $\epsilon_{ij}S_iS_j=1$. Therefore, if we identify
the variables $S_i$ with the orientation of a non-symmetrical molecule,
this condition means that two molecules can be near only if their
relative orientation is appropriate.

Being constituted essentially by diffusing particles,
this model is suited to study quantities like
the diffusion coefficient, or the density autocorrelation functions, that
are usually important in the study of liquids.
Indeed, the model has proven to reproduce fairly well many features
of supercooled glass-forming liquids, as for example the ``cage effect''.
At low temperature and high density, the model shows a two step
relaxation in the self correlation function
and in the mean square displacement. Furthermore, being a finite dimensional
model, it is suitable to study activated processes, that are absent in mean 
field models.

Although the model has been extensively studied by Monte Carlo simulations
\cite{ref:filgmc} and in mean field \cite{ref:filgmf}, there are still
many unsolved problems, concerning the type of transition presented by
the model.
In particular, it would be interesting to study if, in finite dimension,
density and spin variables
become critical at the same point, and what kind of replica
symmetry breaking they show in the spin glass phase.

In this paper, we study the static and dynamical equilibrium properties of
the model in the limit $J\to\infty$.
We will evaluate the equilibrium overlap distribution of spin and density
variables, the equilibrium autocorrelation functions of the self-overlaps,
and the self-diffusion coefficient.

\section{Spin and density overlap distributions}
We have simulated the frustrated Ising lattice gas
by means of the parallel tempering technique \cite{ref:huku,ref:young}.
With this technique one can thermalize the system at high chemical
potential (high density), where the conventional Monte Carlo algorithms
suffer of extremely long autocorrelation times.
One has to simulate several identical replicas of the system, at
different chemical potentials $\mu_0<\cdots<\mu_n$,
where $\mu_0$ corresponds to a low density and very short autocorrelation
time, and $\mu_n$ to the highest value of the chemical
potential that one wants to investigate.
Each step of the algorithm consists of the following two substeps: for each
replica, perform a conventional Monte Carlo step with the given
temperature and chemical potential; for each pair of replicas with 
adjacent chemical potentials, try to swap them with probability
$P_{\text{swap}}=\min(1,\exp(-\beta\Delta\mu\Delta n))$, where 
$\Delta n$ is the difference in the number of particles of the
two replicas considered, and $\Delta\mu$ the difference of their chemical
potentials.
If one chooses carefully the set of chemical
potentials, then the replicas will perform a random walk over the interval
$[\mu_0,\mu_n]$. The time needed to go from $\mu_0$ to $\mu_n$
and back again is called {\em ergodic time}, and can be considered as the
maximum autocorrelation time of the system.

We have simulated the model for $J=\infty$ and $\beta=1$,
between the chemical potentials $\mu_{\text{min}}=1.69$ and
$\mu_{\text{max}}=10.69$, for system sizes $6^3$, $8^3$ and $10^3$.
In the first case we have performed the simulation over 12 values
of the chemical potential, in the second over 16 values, and in the
third over 25 values. The exact values of the chemical potentials were
determined by an algorithm which tried to optimize the intervals, in order
to obtain the same swap rate between all the adjacent replicas. 
For each value of the chemical potential we simulated
two replicas, in order to evaluate spin and density overlap.
The thermalization of the systems was checked by looking at the densities of
the replicas, and waiting until they did not show any sensible drift in time.
The thermalization time was greater than $10^6$ steps.

After having thermalized the systems,
at each parallel tempering step we collected, for each pair of replicas at the
same chemical potential, the spin overlap
$q_s={1\over N}\sum_iS^\alpha_in^\alpha_iS^\beta_in^\beta_i$
and the density overlap $q_d={1\over N}\sum_in^\alpha_in^\beta_i$,
where the superscripts $\alpha$ and $\beta$ refer to the two replicas,
and $N$ is the number of spins. We simulated the systems for up to $10^7$
steps for the largest size, and
checked the symmetry of the spin overlap distribution to be sure that 
the simulation time was sufficient.
All the relevant quantities were averaged over 32 disorder realizations,
and errors were evaluated from the fluctuations between different
realizations.

\begin{figure}
\begin{center}
\mbox{\epsfysize=8cm\epsfbox{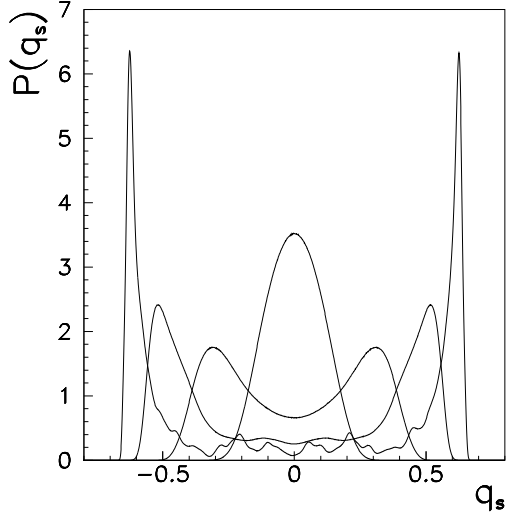}}
\end{center}
\caption{Spin overlap distribution $P(q_s)$ for size $10^3$ and chemical
potentials $\mu$=3.08, 4.12, 5.79, 10.69.}
\label{fig:qs}
\end{figure}
\begin{figure}
\begin{center}
\mbox{\epsfysize=8cm\epsfbox{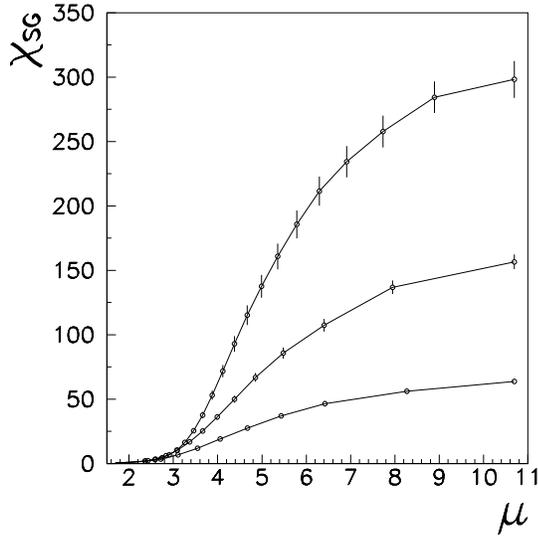}}
\end{center}
\caption{Spin glass susceptibility $\chi_{SG}$ as a function of the chemical
potential, for sizes $6^3$, $8^3$ and $10^3$.}
\label{fig:chisg}
\end{figure}
\begin{figure}
\begin{center}
\mbox{\epsfysize=8cm\epsfbox{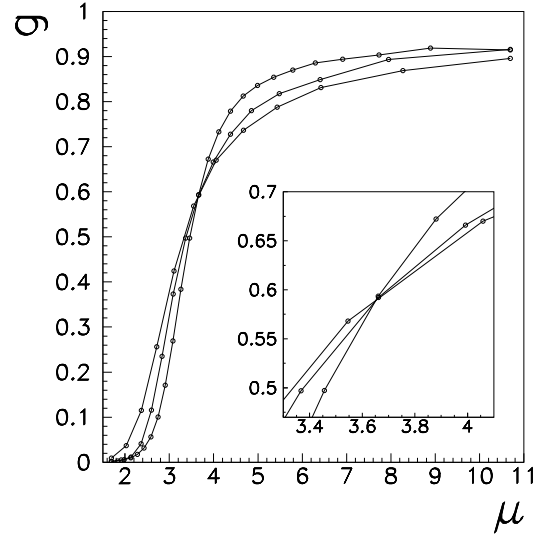}}
\end{center}
\caption{Binder parameter $g$ as a function of the chemical potential,
for sizes $6^3$, $8^3$ and $10^3$. Inset: the point where the curves
cross, at $\mu=3.67$.}
\label{fig:binder}
\end{figure}
\begin{figure}
\begin{center}
\mbox{\epsfysize=8cm\epsfbox{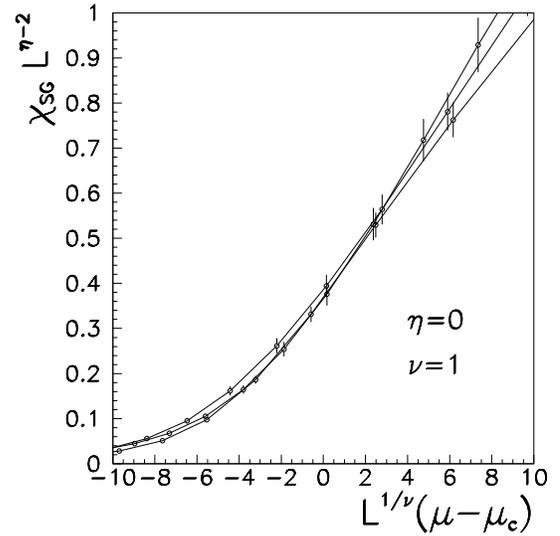}}
\end{center}
\caption{Finite size scaling
plot of the spin glass susceptibility, with $\mu_c=3.67$.
The exponents that give the best
data collapse are $\eta=0$ and $\nu=1$.}
\label{fig:fss}
\end{figure}

In Fig. \ref{fig:qs} the equilibrium distribution of the spin overlap is shown,
for the largest size $10^3$ and for different chemical potentials.
For high chemical potential, the distribution
develops two peaks, separated by a continuous plateau.
This is typical
of models with a continuous replica symmetry breaking, like the Ising
spin glass in three dimensions. Note that, for the highest chemical potential,
the distribution between the two peaks is not a constant plateau,
but is formed
by many small peaks. This can be due to the fact that, at high chemical
potential or low temperature, averaging over more than 32 disorder
realizations is needed to obtain a constant plateau. 
The continuous replica symmetry breaking is associated with the divergence
of the spin glass susceptibility, defined as $\chi_{SG}=N\lan q_s^2\ran$,
where $\lan\cdots\ran$ denotes both the thermal average for a given set of
interactions and the average over the disorder realizations. We have evaluated
$\chi_{SG}$ for the different sizes and chemical potential, and the result
is shown in Fig. \ref{fig:chisg}. The behavior of the susceptibility
confirms the presence of a thermodynamical second order transition. The exact
value of the chemical potential at the transition can be evaluated by looking
at the Binder parameter $g={1\over 2}(3-\lan q_s^4\ran/\lan q_s^2\ran^2)$,
that is 
shown in Fig. \ref{fig:binder}. The curves corresponding to the different 
sizes cross at the chemical potential $\mu_c=3.67$. Once we have
located the transition, we can try to evaluate the critical exponents, using
the relation
$\chi_{SG}(L,\mu)=L^{2-\eta}\tilde\chi_{SG}[L^{1/\nu}(\mu-\mu_c)]$,
that should be valid around the transition, with $\tilde\chi_{SG}[x]$ a
universal curve. In Fig. \ref{fig:fss} the best fit is shown, obtained for
the values $\eta=0$ and $\nu=1$. Note that these exponents were obtained
also by Campellone {\em et al.} \cite{ref:3dps} in the three-dimensional
version of the $p$-spin model. They are different from the exponents
found in the Ising spin glass, $\nu=1.7\pm 0.3$
and $\eta=-0.35\pm 0.05$ \cite{ref:ISG}. This suggests that the transition
could belong to a different universality class with respect to the 
Ising spin glass.

\begin{figure}
\begin{center}
\mbox{\epsfysize=8cm\epsfbox{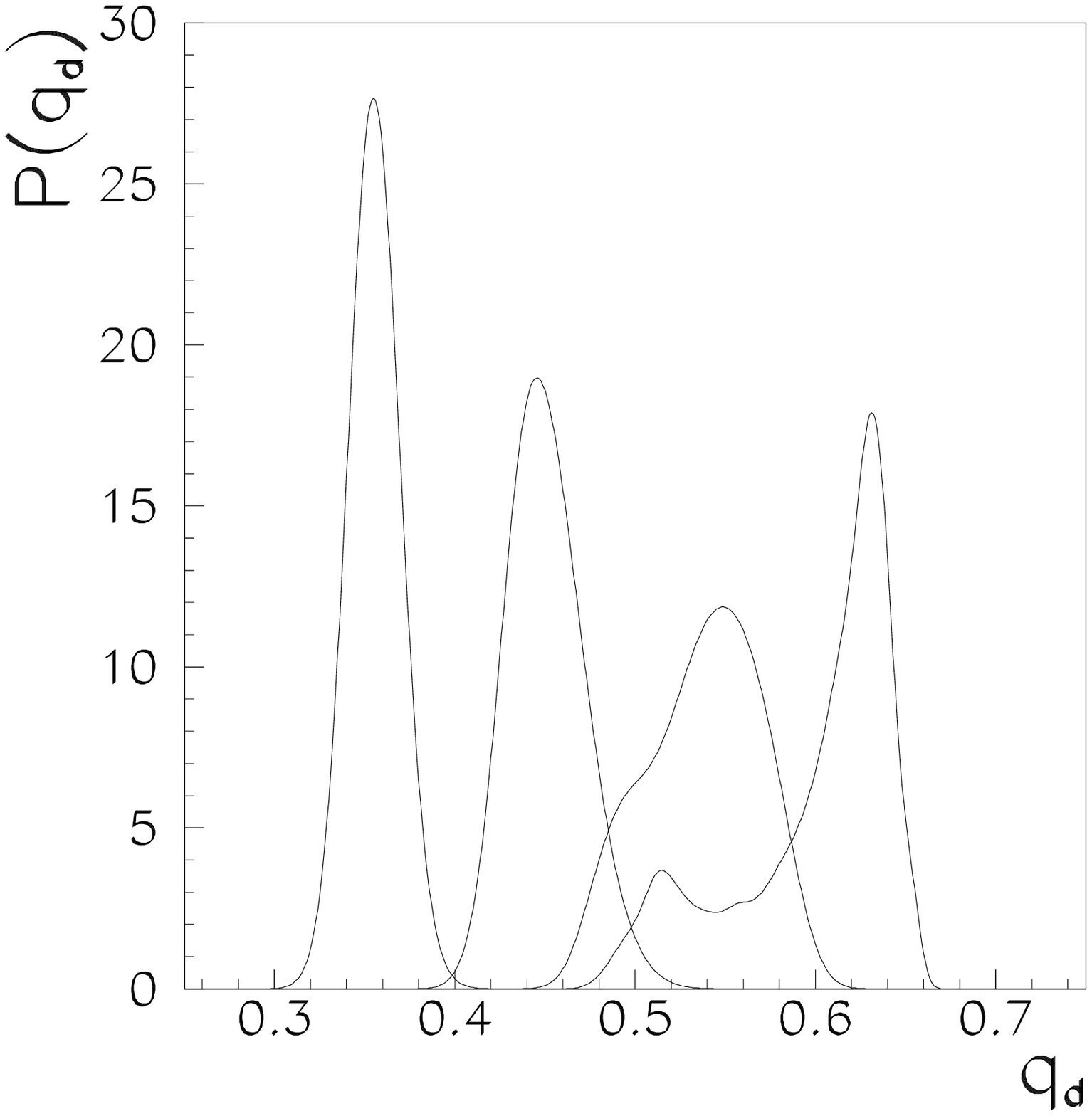}}
\end{center}
\caption{Density overlap distribution $P(q_d)$ for size $10^3$ and
chemical potentials $\mu$=3.08, 4.12, 5.79, 10.69.}
\label{fig:qd}
\end{figure}

\begin{figure}
\begin{center}
\mbox{\epsfysize=8cm\epsfbox{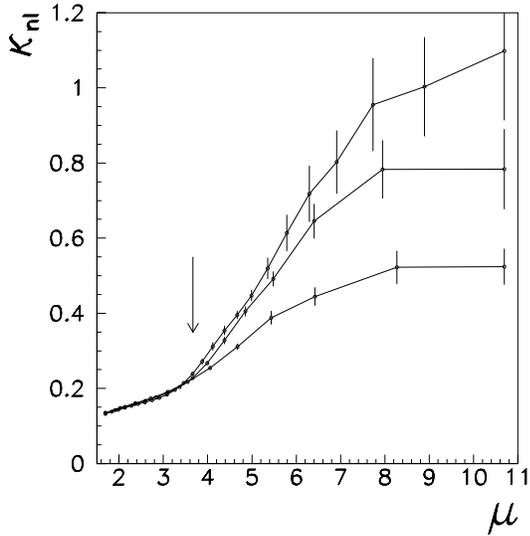}}
\end{center}
\caption{Non-linear compressibility $\kappa_{nl}$ as a function of
the chemical potential, for sizes $6^3$, $8^3$ and $10^3$. The arrow
marks the point where spin variables display the transition.}
\label{fig:chin}
\end{figure}

We have then evaluated the equilibrium distribution of the density
overlap, which is shown in Fig. \ref{fig:qd} for the largest size and for
different chemical potentials. The non-linear compressibility 
$\kappa_{nl}=N(\lan q_d^2\ran-\lan q_d\ran^2)$ is shown in Fig.
\ref{fig:chin} as a function of the chemical
potential. The arrow in  Fig. \ref{fig:chin} marks the point where the
spin variables undergo the transition, as signaled by the crossing of the
Binder parameter, and the spin glass susceptibility diverges. It is
evident that this point does not correspond to a divergence of the 
non-linear compressibility. Therefore the divergence of the spin glass
susceptibility is due to the fluctuations of spin variables $S_i$,
and not to the fluctuations of the density variables $n_i$. 
We have also evaluated the Binder parameter of the density overlap (not
shown), which shows a non-monotonic
behavior, becoming negative for low density and positive at high density,
similarly to what is observed in the finite-dimensional p-spin
\cite{ref:3dps}. Indeed the curves for different sizes do not cross at a 
definite point, so they cannot be used to localize the transition (if any)
of the density variables.

It remains to be determined whether or not 
the density variables exhibit a transition
not manifested by a divergence of the non-linear compressibility.
Note that for very high chemical potential the equilibrium distribution of
the density overlap develops a secondary minimum.
This could correspond
to a transition of different kind, perhaps similar to the 1-step
replica symmetry breaking transition of the $p$-spin models.

\section{Relaxation functions of spin and density self-overlap}
The frustrated Ising lattice gas is known to have very large relaxation
times at high density or low temperature \cite{ref:filgmc}. Here we want
to evaluate the relaxation times of the spin self-overlap, defined as
$q_s(t)={1\over N}\sum_i\lan S_i(\tp)n_i(\tp)S_i(\tp+t)n_i(\tp+t)\ran$,
and the density self-overlap
$q_d(t)={1\over N}\sum_i(\lan n_i(\tp)n_i(\tp+t)\ran-\lan n_i(\tp)\ran^2)$,
where the average $\lan\cdots\ran$ is done over the reference time $\tp$.
We have simulated the model for system size $20^3$, and 10
chemical potentials between $\mu=2.583$ and 3.661, in the following 
manner. We start with an empty system, with the interactions randomly 
chosen, thermalize at the given chemical potential for a given time
$\Delta t$, save the obtained configuration, and then simulate the model
saving the self-overlaps $q_s(t)$
and $q_d(t)$ with respect to the configuration at the end of thermalization.
Then we repeat the process again with a different disorder configuration
and thermal noise.
The thermalization time $\Delta t$ is at least 10 times larger than 
the time needed to the self-overlap to decay to the value 0.1, except for the
three highest chemical potentials, for which the thermalization time was
shorter. The self-overlaps were averaged over at least 100 different runs,
and errors were evaluated from the fluctuations between different runs.
When evaluating the density self-overlap $q_d(t)$, the quantity 
${1\over N}\sum_i\lan n_i(\tp)\ran^2$ can be taken equal to the average
density overlap $\lan q_d\ran$ at equilibrium, as calculated in the 
parallel tempering simulations.

\begin{figure}
\begin{center}
\mbox{\epsfysize=8cm\epsfbox{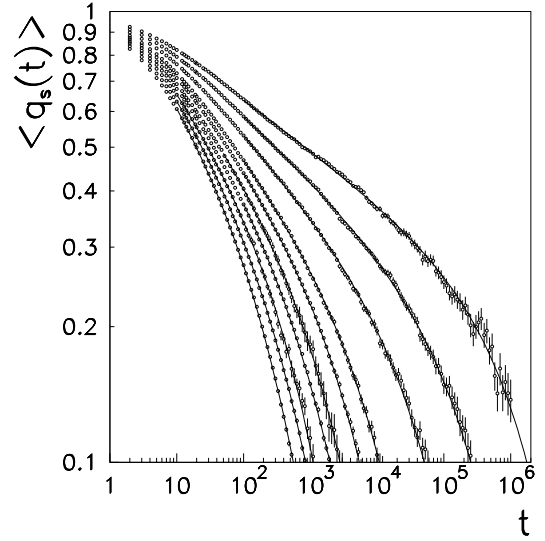}}
\end{center}
\caption{Relaxation functions of the spin self-overlap, for system size $20^3$
and chemical potentials $\mu=2.583$, 2.665, 2.747, 2.829, 2.911, 2.997, 3.083,
3.264, 3.456, 3.661. Continuous lines are fits with the function
$ct^{-x}\exp(-(t/\tau)^\beta)$.}
\label{fig:spin}
\end{figure}

\begin{figure}
\begin{center}
\mbox{\epsfysize=8cm\epsfbox{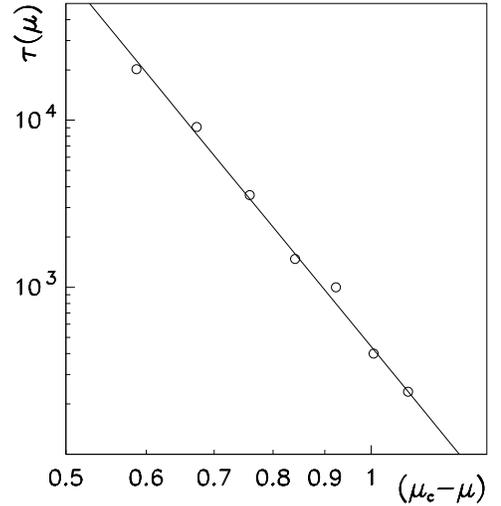}}
\end{center}
\caption{Relaxation times $\tau$ of the spin self-overlap, as obtained
by the fits of Fig. \ref{fig:spin}, for chemical potentials
$2.583\le\mu\le 3.083$.
The straight line is a fit with the function $|\mu-\mu_c|^{-z\nu}$, and
$\mu_c=3.67$ fixed, which gives $z\nu=7.4$.}
\label{fig:tau}
\end{figure}

In Fig. \ref{fig:spin} the relaxation functions of the spin self-overlap
are shown. The solid lines are fit with the function 
$ct^{-x}\exp(-(t/\tau)^\beta)$, proposed by Ogielski for the 
Ising spin glass \cite{ref:ogi}. The exponent $\beta$ is nearly constant
within the errors for all the chemical potentials considered, and
slightly greater than 0.5, while the exponent $x$ varies between 0.2
for the lowest chemical potential to 0.1 for the highest.
The correlation times
$\tau$ are shown in Fig. \ref{fig:tau}, excluding the last three points,
which are likely to suffer from insufficient thermalization or finite
size effects. A power law fit
$\tau\sim|\mu-\mu_c|^{-z\nu}$ gives a 
dynamical exponent $z=7.4$, slightly 
greater than the exponent $z=6.0\pm 0.8$ found by Ogielski for the Ising
spin glass.

\begin{figure}
\begin{center}
\mbox{\epsfysize=8cm\epsfbox{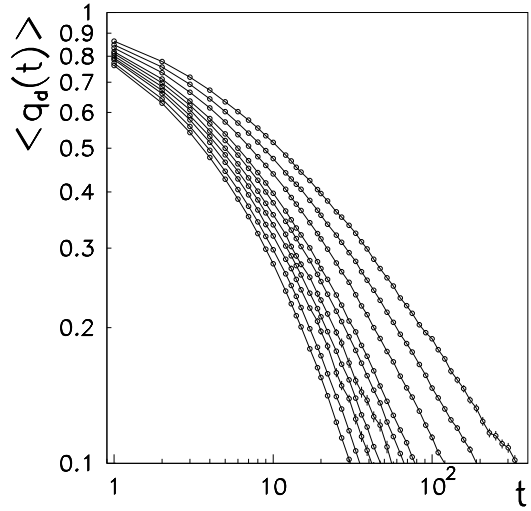}}
\end{center}
\caption{Relaxation functions of the density self-overlap, for the same
system size and chemical potentials of Fig. \ref{fig:spin}.}
\label{fig:rho}
\end{figure}

In Fig. \ref{fig:rho} the relaxation functions of the density
self-overlap are shown, for the same system size and chemical 
potentials. Note that the relaxation times grow very slowly with respect to
those of the spin self-overlap. This supports the conclusion that 
the transition at $\mu_c=3.67$ does not involve the density variables $n_i$.
The latter probably undergo a transition at a higher chemical potential,
inside the phase where the spin variables are frozen.

\section{Diffusivity}
We have simulated the model with a purely diffusive dynamics, in the following
way. We start with an empty lattice, with random interactions, and 
then slowly raise the chemical potential, until a given density is reached.
Then we switch to a purely diffusive dynamics, with conserved number of
particles, and thermalize the system at the given density. After having 
thermalized the system, we collect the mean square displacement
$\lan r^2(t)\ran$ of the 
particles as a function of time. The long time regime of the mean square
displacement is of the form $\lan r^2(t)\ran=Dt$, from which we extract
the diffusion coefficient $D$. To each density, we associate a chemical
potential from the equilibrium relation between the two quantities.

\begin{figure}
\begin{center}
\mbox{\epsfysize=8cm\epsfbox{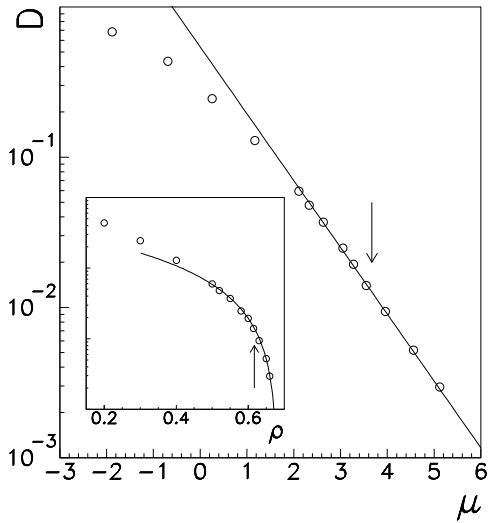}}
\end{center}
\caption{Diffusivity as a function of the chemical potential,
for a system size $16^3$. The solid line is a fit with the
Arrhenius form $D=ae^{-\alpha\mu}$. Inset: Diffusivity as a function
of the density. The solid line is a fit with the power law
$D=a(\rho_0-\rho)^\gamma$, with $\rho_0=0.681$ and $\gamma=1.38$.
The arrows mark the point where the spin variables display the transition.}
\label{fig:diff}
\end{figure}

In Fig. \ref{fig:diff} the diffusion coefficient $D$ is shown for
a system size $16^3$ as a function of the chemical potential. For high 
chemical potential, it can be well fitted by an Arrhenius form, 
$D=ae^{-\alpha\mu}$. Therefore, the diffusion of the particles seems
to stop only at $\mu\to\infty$, which corresponds to $T\to 0$. The arrow
marks the point where the spin variables undergo the spin
glass-like transition: apparently no anomaly in the diffusivity shows up
in correspondence of the transition. In the inset, the diffusivity as
a function of the density is shown. Note that for $\mu\to\infty$
the density goes to a maximum value $\rho_{\text{max}}\simeq 0.68$.
The diffusivity can be well fitted by a power law
$D=a(\rho_0-\rho)^\gamma$, where $\rho_0=0.681$ and $\gamma=1.38$.

\section{Conclusions}
We have studied the static and dynamical properties of the frustrated
Ising lattice gas at equilibrium. 
A spin glass-like transition is found in the spin variables, signaled by the
crossing of the Binder parameter, the divergence of the non-linear
susceptibility, and the development of a continuous replica symmetry breaking
in the spin overlap distribution. The equilibrium autocorrelation
functions of the spin overlap can be well fitted by an Ogielsky form, with
a correlation time diverging at the critical point.
On the other hand, the density variables seem to be affected little by the
transition, showing no divergence either in the non-linear compressibility,
or in the autocorrelation time.

The freezing of the model at the chemical potential $\mu_c$ is therefore
connected with a second order transition in the spin variables, more similar
to the freezing of the Ising spin glass than to the mode-coupling transition
of structural glasses. One cannot exclude that the density variables undergo
a $p$-spin-like transition at a higher density, characterized by a 1-step
replica symmetry breaking and a discontinuity of the Edwards-Anderson
parameter defined in terms of density variables. This fact is suggested 
by the development of a secondary peak in the density overlap distribution
at very high chemical potential, as well as by the measurements of 
the off-equilibrium fluctuation-dissipation ratio \cite{ref:fdr},
but more work is needed to clarify this point.
\section*{Acknowledgments}
This work was partially supported by the European TMR Network-Fractals
(Contract No. FMRXCT980\-183), MURST-PRIN-2000 and INFMPRA(HOP).
We acknowledge the
allocation of computer resources from INFM Progetto Calcolo Parallelo.
%
%
%

%
%
%
%
\end{document}